\documentstyle{mn}
\input{epsf}
\newcommand{\be}{\begin{eqnarray}}
\newcommand{\ee}{\end{eqnarray}}
\newcommand{\hMpc}{{\ifmmode{h^{-1}{\rm Mpc}}
\else{$h^{-1}$Mpc}\fi}}

\title[Evolution of the DM LSS ]
{Simulated evolution of the dark matter large-scale structure}
\author[Demia\'nski,  Doroshkevich, Pilipenko, \& Gottl\"ober]
        {M. Demia\'nski$^{1,2}$, A. Doroshkevich$^{3}$,
        S. Pilipenko$^{3}$, S. Gottl\"ober$^4$\\
        $1$Institute of Theoretical Physics,
                       University of Warsaw,
                       00-681 Warsaw, Poland\\
        $2$Department of Astronomy, Williams College,
           Williamstown, MA 01267, USA\\
        $3$Astro Space Center of Lebedev Physical
           Institute of  Russian Academy of Sciences,
                        117997 Moscow,  Russia\\
        $4$ Astrophysical Institute Potsdam,
            An der  Sternwarte 16, 14482 Potsdam, Germany \\
}

\date{Accepted 25.11.06,
      Received ...,
        in original form ... .}

\begin{document}
\maketitle

\begin{abstract}
We analyze evolution of the basic properties of simulated large
scale structure elements formed by dark matter (DM LSS) and confront
it with the observed evolution of the Lyman-$\alpha$ forest. In
three high resolution simulations we selected samples of compact DM
clouds of  moderate overdensity. Clouds are selected at redshifts
$0\leq z\leq 3$ with the Minimal Spanning Tree (MST) technique.  The
main properties of so selected clouds are analyzed in 3D space and
with the core sampling approach, what allows us to compare estimates
of the DM LSS evolution obtained with two different techniques and
to clarify some important aspects of the LSS evolution. In both
cases we find that regular redshift variations of the mean
characteristics of the DM LSS are accompanied only by small
variations of their PDFs, what indicates the self similar character
of the DM LSS evolution. The high degree of relaxation of DM
particles compressed within the LSS is found along the shortest
principal axis of clouds. We see that the internal structure of
selected clouds depends upon the mass resolution and scale of
perturbations achieved in simulations. It is found that the low mass
tail of the PDFs of the LSS characteristics depends upon the
procedure of clouds selection.
\end{abstract}


\section{Introduction}

At $z\leq 1.5$ the Large Scale Structure of the Universe (LSS) is
observed in many galaxy catalogs such as 2dFGRS, SDSS, DEEP2, VIRMOS
(Percival et al, 2001; Verde et al. 2003; Abazajian et~al. 2003;
Davis et al. 2003; Le Fevre et al. 2005). It is manifested as a
strong galaxy concentration within rich walls and filaments
surrounding regions with low density of galaxies (voids). At
redshifts $z\geq 2$ the LSS is observed as the Ly-$\alpha$ forest
and rare metal systems in high resolution spectra of the farthest
quasars. The observed properties of these populations of the LSS
elements are different but now it is commonly believed that both of
them represent {\it different manifestations} of the same structure
formed by DM and baryonic components. This statement suggests that
the Ly-$\alpha$ absorbers can be associated with low mass structure
elements formed by the non luminous baryonic and DM components,
while galactic walls and filaments represent fraction of the richer
LSS elements. This means that the DM structure traced by the
Ly-$\alpha$ absorbers is qualitatively similar to the rescaled structure
observed in the spatial distribution of galaxies. In particular,
this inference is consistent with observations of absorbers in the
vicinity of galactic filaments and even galaxies within voids
(Morris et al. 1993a, b; Penton, Stock and Shull 2000, 2002; McLin
et al. 2002, Williger et al. 2010).

The facilities of both observational approaches are limited
but they are complementary to each other. Thus analysis of
the galaxy surveys is focused on the richer LSS elements
at limited range of redshifts $z\leq 1 - 1.5$. It allows us to
determine the main characteristics of such elements ( see,
e.g., Doroshkevich et al. 2004) but cannot trace the LSS
evolution for large redshift intervals. This analysis is
concentrated on the investigations of galactic properties
and their environmental
dependence. For example, it allows us to test the star formation
histories (see, e.g., Panter et al. 2007; Skibba et al. 2009)
and correlations between orientations of the LSS elements and
the angular momentum of galaxies (see, e.g., Trujillo et al.
2006; Aragon--Calvo et al. 2007; Paz et al. 2008; Slozar et
al. 2009; Jimenez 2009).

On the other hand, observations of the Ly-$\alpha$ forest at $z\sim
2 - 4$ allow us to trace some characteristics of the low mass LSS
elements and their evolution. However, they rely on the observed
evolution of neutral hydrogen, which can be caused by many factors.
Thus, in addition to the actual evolution of the DM components of
the Ly-$\alpha$ absorbers the observed fraction of neutral hydrogen
depends upon poorly known variations of the UV background. In spite
of this, it is clear that there is a close link between the observed
Ly-$\alpha$ absorbers and DM clouds formed at high redshifts.

It seems that the close link between the Ly-$\alpha$ forest and the
galaxy LSS can be established by numerical simulations of the
structure evolution. However, it is not yet possible to simulate
the LSS evolution in a wide range of scales. Thus, simulations
performed within large boxes with a moderate spatial and mass
resolution (for review see Frenk 2002; Springel, Frenk\,\&\,White
2006) reproduce reliably only the formation of the richer DM LSS
elements similar
to those observed in the galaxy distribution. On the other hand,
high resolution simulations performed with the small box sizes can
reproduce some characteristics of the Ly-$\alpha$ forest (see, e.g., Weinberg et
al. 1998; Zhang et al. 1998; Dav\'e et al. 1999; Theuns et al. 1999,
2000; Schaye 2001; Meiksin, Bryan and Machacek 2001) but obviously
they cannot simulate formation of the richer LSS elements. Perhaps,
more progress can be achieved with simulations performed with variable
resolution (see, e.g., Springel et al. 2008; Diemand et al.
2008).

Technical limitations of the box size and resolution lead to small
and large scale cutoffs of the simulated power spectrum. These
cutoffs restrict the potential of such simulations and do not allow
to reproduce reasonably well the observed characteristics of the
Ly-$\alpha$ forest (see, e.g., Meiksin, Bryan and Machacek 2001;
Gnedin \& Hamilton 2002; Tegmark \& Zaldarriaga 2003; Seljak,
McDonald \& Makarov 2003; Manning 2003a,b; Demia\'nski \&
Doroshkevich 2003). Now such simulations are used mostly for the
surprisingly stable reconstruction of the small scale initial power
spectrum from characteristics  of the forest (see, e.g., Croft et al.
2001, 2002; Seljak et al. 2005; McDonald et al. 2005; Viel et al
2004a, b).

An alternative approach is to focus the main attention on the regular
trends in the evolution of the simulated LSS and compare them
with observations of the forest properties in a wide range of
redshifts. Such trends can be revealed with special methods
applied to representative high resolution simulations. The
successful investigation of the process of halo formation (see,
e.g., Navarro, Frenk, \& White, 1995, 1996, 1997; Bullock et al.,
2001; Tasitsiomi et al. 2004) demonstrates the high potential of
this approach.

The statistical analysis of a large sample of the Ly-$\alpha$
absorbers (Demia\'nski, Doroshkevich \& Turchaninov 2006;
Demia\'nski \& Doroshkevich 2010) reveals some unexpected features
of the forest evolution. The most important are the regular redshift
variations of the mean observed characteristics of the Ly-$\alpha$
absorbers and the surprisingly weak variations of their probability
distribution functions (PDFs). These results demonstrate the self
similar character of evolution of the observed LSS, which must be
tested with numerical simulations.

Other important open problems are connected with the internal
structure of the Ly-$\alpha$ absorbers and its evolution with
time. They also include an estimate
of the degree of relaxation of matter accumulated by absorbers
with various richness, and the link between the thermal and
large scale bulk motions of the gas and the observed
Doppler parameters of the forest. At the present time a reliable
observational discrimination between contributions of these
factors is problematic. We can also note the surprisingly
weak redshift evolution of the observed mean Doppler parameter
and a complex shape of its PDF indicating the existence of a rich
sample of DM clouds with small Doppler parameter $b\leq \langle
b\rangle$.

Comparison of the observed evolution of the forest with the
simulated evolution of the DM clouds allows us to clarify how
they are related. For this purpose, in this paper, we investigate
evolution of the DM LSS in high resolution simulations using
the Minimal Spanning Tree (MST) and Core Sampling approaches
(see, e.g., Doroshkevich et al. 2004).
The space and force resolutions achieved in these
simulations allow us to characterize the evolution of the
compact DM LSS elements with moderate richness, which is by
itself an important problem. Thus, we reveal significant
differences in properties and evolution rates of the
LSS elements selected in simulations with different mass and
space resolutions. These results demonstrate the strong
influence of small scale perturbations on the internal structure
of the LSS elements.

All selected DM elements are significantly larger than those
observed as the Ly--$\alpha$ forest, what does not allow us
to perform a
direct quantitative comparison of the observed and simulated
clouds. However, the basic properties of the selected sample
of DM clouds are found to be similar to the properties of
the observed Ly-$\alpha$ forest. In
particular, we confirm the self similar character of
evolution of the DM LSS, reproduce the redshift dependence
and the PDF of the Doppler parameter, estimate the clouds
rotation, the contribution of macroscopic turbulent motions
and the degree of relaxation of compressed matter measured
along the principal axes of clouds. The properties of the LSS
elements are found to be qualitatively consistent with
predictions of the Zel'dovich approximation (Zel'dovich 1970;
Shandarin \& Zel'dovich 1989; Demia\'nski\,\&\,Doroshkevich
1999, 2004).

This paper is organized as follows.
Simulations that we use  and methods of their analysis
are shortly presented in Sec. 2. Evolution of the basic
characteristics of the DM LSS is described in Sec. 3 and
results of the core sampling analysis are given in Sec. 4.
Discussion of obtained results and conclusions can be found
in Sec. 5.

\section{Numerical simulations}

In this paper we investigate the process of formation and evolution
of the DM LSS within the concordance cosmological model
\[
\Omega_\Lambda=0.7,\quad \Omega_m=0.3,\quad h=0.7,\quad
\sigma_8=0.9,\quad n=1\,,
\]
where $\Omega_\Lambda$ and $\Omega_m$ denote the dimensionless
densities of Dark Energy and matter, $h=H_0/100km/s/Mpc$  the
dimensionless Hubble parameter, $\sigma_8$ - the amplitude of
density perturbations, $n$ - the power index of the spectrum of
perturbations.

One simulation - referred further as $S_{150}$ -
was performed with an MPI version of the Adaptive Refinement
Tree code (Kravtsov et al. 1997) within a box of $L_{box}=
150h^{-1}$Mpc with 256$^3$ particles, cell size $\approx
0.6h^{-1}$Mpc and mass and force resolutions of $1.7\cdot
10^{10}h^{-1}M_\odot$ and 18$h^{-1}$kpc (see Wojtak et al.,
2005 for more details). We have analyzed the simulated DM
distribution at four redshifts, namely, at $z=0, 1, 2\,\&\,3$,
what roughly covers the interval of observed redshifts.

For comparison we also used the DM distribution obtained in the
MareNostrum Universe (Gottl\"ober, Yepes 2007). This non-radiative
SPH simulation was performed with GADGET2 code (Springel 2005) using
the concordance model.  This simulation - further referred as
$S_{500}$ - consists of $1024^3$ dark and $1024^3$ gas particles in
a box of $L_{box}=500\;h^{-1}$Mpc on a side. Both the mass and force
resolutions ($0.75\cdot 10^{10}h^{-1}M_\odot,\,\,15\;h^{-1}$
comoving kpc) are comparable to those achieved in the previous
simulation. Such comparison allows us to check the impact of the
baryonic component, the code used and the initial realization on the
properties of the simulated LSS. As a rule, properties of clouds
selected in this simulation are close to the ones obtained for the
simulation $S_{150}$, what demonstrates the weak impact of these
factors on the characteristics of the LSS.

The third simulation  - further referred as $S_{50}$ - consists of
$512^3$ DM particles in a box of $L_{box}=50\;h^{-1}$Mpc on a side.
It was performed at the Astrophysical Institute Potsdam by M.
Steinmetz. With such parameters the obtained mass resolution is
$7.8\cdot 10^7h^{-1}M_\odot$ and the force resolution $3h^{-1}$
comoving kpc, which is better than resolutions achieved in the
previous simulations. However, in this case the important influence
of large scale perturbations is partly suppressed. Comparison of
results obtained for these three simulations reveals the impact of
small scale perturbations on the properties of the LSS. In
particular, in this simulation we see more complex internal
structure of the selected LSS elements, earlier formation of high
density clouds, and we can trace the initial steps of the LSS
formation up to redshifts $z\sim 12 - 14$.

\subsection{Selection of high density clouds}

Using the Minimal Spanning Tree (MST) code as described in
Doroshkevich et al. (2004) we select from the full distribution of
DM particles a set of clouds with three threshold  parameters that
restrict their density and richness. They are the threshold
overdensity, $\delta_{thr}$, and the minimal and maximal richness,
$N_{min}\,\&\,N_{max}$, of selected clouds \be \delta_{thr}= (4\pi
\ell_{thr}^3\langle n_p \rangle/3)^{-1}, \quad N_{min}\leq N_p\leq
N_{max}\,, \label{dthr} \ee
\[
\langle n_p\rangle=(5,\,8.6\,\&\,10^3)h^3{\rm Mpc}^{-3},
~~ \langle\rho\rangle\approx 8\cdot 10^{10}h^2M_\odot/Mpc^3\,.
\]
Here $\ell_{thr}$ is the maximal length of the edges of the tree
within a selected cloud, $\langle\rho\rangle$ and $\langle
n_p\rangle$ are the mean comoving density and number density of DM
particles in the simulations. Using these parameters we can reliably
identify clouds even of a complex irregular shape. By definition of
the MST the distances between all neighboring points inside the
cloud are less than the threshold one, $\ell\leq\ell_{thr}$, and,
therefore, $\delta =\rho/\langle\rho\rangle\geq\delta_{thr}$. In
other words, the selected clouds are bound by the surfaces of
constant overdensity, $\delta= \delta_{thr}$.

In this paper we are mainly interested in the mildly nonlinear
evolution of the LSS elements with moderate overdensity. Such
elements dominate at the early period of the LSS formation and
it can be expected that we observe them as the Ly-$\alpha$
forest. Hence the threshold overdensities that we adopt in this
paper are not very high: $\delta_{thr}=1$ and $\delta_{thr}
=1.76$.

The other important threshold parameters are the minimal and maximal
richness of the selected clouds. The first one ensures the
reliability and stability of characteristics of the selected clouds,
while the second one allows us to exclude from the analysis
amorphous multiconnected clouds formed at later stages of the LSS
evolution. Such extremely rich clouds appear owing to the
integration of less rich clouds and they represent the elements of
the well known network of the LSS. Evidently, such clouds can not be
described by just a few local characteristics.

For these simulations we use the following minimal and maximal
richness of the clouds under investigation
\[
N_{min}=30,\quad \,\,\,\,\,\,\,\, N_{max}=5000,\quad {\rm for
\,\,S_{150}}\,,
\]
\be
N_{min}=70,\quad  \,\,\,\,\,\,\,N_{max}=11\,000,\quad {\rm for
\,\,S_{500}}\,,
\label{nn}
\ee
\[
N_{min}=6.5\cdot 10^3,\quad N_{max}=10^6\quad {\rm for
\,\,S_{50}}\,,
\]
what approximately corresponds to the same range of cloud masses,
\be
M_{min}\simeq 5\cdot 10^{11}h^{-1}M_\odot,\quad M_{max}\simeq 8
\cdot 10^{13}h^{-1}M_\odot\,.
\label{nm}
\ee

Further on we refer to the samples selected with $M_{min}\leq M_{cl}
\leq M_{max}$ as samples of compact clouds. For some tests
we also use samples with $M_{cl}>M_{min}$ (a full sample) and
$M_{cl}>M_{max}$ (a sample of rich clouds).

The selection of clouds with the core sampling approach is
described in Sec. 4.

\subsection{Characteristics of clouds}

In this paper clouds are characterized by their principal axes
determined by their inertia tensor. Such rough description allows us
to take into account the main global shape of the clouds and to
estimate their degree of filamentarity and sheetness without
introduction of any additional parameters.

Sometimes much more refined techniques are used for selection and
discrimination of the DM filaments and walls (see, e.g., discussion
in Doroshkevich et al. 2004; Aragon--Calvo et al. 2007; Zhang et al.
2009). However, advantages of such approach are problematic owing to
very complex shapes, many branches and strong nonhomogeneity of the
matter distribution typical for both the observed and simulated LSS
elements. In contrast to our approach such refined methods are
inevitably multiparametric and therefore description of so selected
clouds becomes more complicated. Such approach can be efficient in
providing more accurate description of local environment of
considered objects.

Here we consider several global characteristics of clouds,
namely, their mass (richness) and velocity, $M_{cl}\,\& \,{\bf U}$,
three components of their angular momenta $J_i$ and the comoving
sizes along the three principal axes, $L_1 \geq L_2\geq L_3$, determined
through their inertia tensors.

As is well known (see, e.g, review by Sh\"afer 2008) the angular
momenta of the clouds are mainly generated by the tidal interactions
with the surrounding medium (Tidal Torque Theory (TTT) Peebles 1969;
Doroshkevich 1970; White 1984). The angular momentum of a cloud depends
upon its (random) shape. The mean angular momentum of clouds is small,
$\langle J_i\rangle\sim 0$ owing to its random orientation and
$\langle J_i^2\rangle$ provides more stable characteristics of
clouds. For a regular ellipsoidal volume with halfaxes $a_1\geq
a_2\geq a_3$, the Zel'dovich theory gives \be \langle
J_i\rangle=0,\quad\langle J_i^2\rangle\propto (a_j^2-a_k^2)^2,\quad
i\neq j\neq k\,, \label{Jth} \ee and, so, $\langle J_2^2\rangle\geq
\langle J^2_{1,3}\rangle $ (Demia\'nski\,\&\,Doroshkevich 2004).

In simulations the angular momenta of clouds and the contribution
of bulk (turbulent) motions are characterized by the functions
\be
J_i=N_p^{-1}\sum_{m=1}^{N_p}\psi_i,\quad \psi_i=\epsilon_{ikl}
({\bf
{v}_m}-{\bf{v}_c})_k({\bf{x}_m}-{\bf{x}_c})_l\,,
\label{turb}
\ee
where $\epsilon_{ikl}$ is the totally antisymmetric unit tensor
and $N_p$ is the number of particles in the cloud.
Here $\bf {x}_m\,\&\,\bf {v}_m$ and $\bf {x}_c\,\&\,\bf {v}_c$ are
the coordinates and velocities of the $m^{th}$ particle and the
center of mass, respectively.

To characterize the internal dynamics of matter accumulated by a
cloud and its degree of relaxation  along the principal axes we
consider also three velocity dispersions, $\sigma_i$, and three
angular momenta of particles, $j_i$,
\be
j_i=N_p^{-1}\sum_{m=1}^{N_p}|\psi_i|\,.
\label{jturb}
\ee
For a
regular ellipsoidal volume with halfaxes $a_1\geq a_2\geq a_3$, the
Zel'dovich theory gives in contrast to (\ref{Jth})
\be \langle
j_i^2\rangle\propto (a_k^2+a_l^2)^2,\quad i \neq k\neq l\,.
\label{zturb} \ee

As is well known, within anisotropic clouds of collisionless
particles the relaxation occurs independently along each
principal axis of the cloud. For example, the observed galactic
walls and filaments are almost relaxed along their shortest
axes but have no time to relax along their longest axis. Thus,
here we will characterize the degree of relaxation of $N_p$
particles compressed within a cloud by the parameters
\be
w_i=1+{1\over N_p\sigma_i}\sum_{m=1}^{N_p}({\bf {v}_m}-
{\bf{v}_c})_i
({\bf{x}_m}-{\bf{x}_c})_i/|({\bf{x}_m}-{\bf{x}_c})_i|\,.
\label{drelax}
\ee
Here $\sigma_i$ is the velocity dispersion and index $i$
shows that we consider the relaxation along the $i^{th}$
principal axis.

Evidently, the functions $w_i$ discriminate between contributions of
random and regular motions. Indeed, for a cloud with relaxed matter
distribution and random velocities of particles we can expect that
$|w_i-1| \ll 1$, because for each particle the probabilities to find
$({\bf v}_m-{\bf v}_c) ({\bf x}_m-{\bf x}_c)_i\leq 0$ and $({\bf
v}_m-{\bf v}_c) ({\bf x}_m-{\bf x}_c)_i\geq 0$ are close to each
other. In contrast, for regular expansion or compression of matter
$({\bf v}_m- {\bf v}_c)$ and $({\bf x}_m-{\bf x}_c)$ are correlated
and, so, $|w_i-1|\sim 1$ seems to be much more probable. The case
when $w_i\geq 1$ or $w_i\leq 1$ corresponds to domination of the
expansion or compression of matter along the $i^{th}$ axis.

The core sampling approach was applied only for the first simulation
and is described in Sec. 4\,.

\subsection{Comparison of the simulations with the Zel'dovich
approximation}

In this paper we consider properties of DM clouds selected in the
same manner in three simulation. The simulations are performed with
different codes, in different boxes and with different resolution.
The cloud selection is performed with a complex algorithm and with
set of thresholds  presented in Sec. 2.1. Therefore before we start our
analysis we estimate the representativity and reliability of the
samples of selected clouds. For this purpose we compare some
integral characteristics of selected clouds with predictions of the
Zel'dovich approximation (Zel'dovich 1970; Shandarin \& Zel'dovich
1989; Demia\'nski \& Doroshkevich 1999, 2004), which reliably
describes the mildly nonlinear evolution of perturbations.

For this aim we use here three characteristics of samples of clouds,
namely, the matter fraction accumulated by the LSS, $f(M)$, the
PDFs of masses of selected clouds, $P(M)$, and the redshift evolution
of the velocity of clouds as a whole, ${\bf U}(z)$. The last test is
more sensitive to the large scale perturbations, while the first and
the second ones are sensitive to the small scale perturbations,
nonlinear processes of matter condensation within clouds and the
procedure of clouds selection.

The shape of PDFs for clouds sizes and internal velocities can be
also obtained on the basis of Zel'dovich theory. However, here we
note only that for these functions and the CDM --like power spectrum
the Gaussian PDF of initial perturbations is transformed to the
exponential one (see sections 2.4 \& 3.3 in Demia\'nski \&
Doroshkevich (2004) for more details). The exponential shape of
these PDFs in our simulations is clearly seen in Figs 1 and 2. In
Sec. 3.3 we compare predictions of the Zel'dovich theory with the
simulated angular momentum and turbulent motions (\ref{Jth},\,
\ref{zturb}).

\subsubsection{Evolution of the cloud velocity}

The nonlinear processes of matter clustering and relaxation of the
compressed matter strongly distort velocities of particles within
clouds but only weakly change the velocities of clouds as a whole.
As noted in Demia\'nski \& Doroshkevich (1999), for the Gaussian
initial perturbations, the PDF of clouds velocity remains Gaussian
with the velocity dispersion closely allied to that described by the
linear theory. Thus, according to the Zel'dovich approximation the
velocity dispersion along any principal direction is \be
\sigma_U={H(z)\over 1+z}\beta(z)B(z){\sigma_s\over \sqrt{3}},\quad
\beta(z)=-{1+z\over B(z)}{dB(z)\over dz}\,, \label{sigu} \ee
\[
B^{-3}\approx 1+{2.2\Omega_m\over 1+1.2\Omega_m}[(1+z)^3-1]\,,
\]
\[
H^2=H_0^2(\Omega_m(1+z)^3+\Omega_\Lambda),
\quad \sigma_s^2={1\over 2\pi}\int_0^\infty p(k)dk\,.
\]
Here $H$ is the Hubble parameter, $H_0=100h$km/s/Mpc, $B(z)$
describes the growth of perturbations in the linear theory for
$\Lambda$CDM cosmology, $p(k)$ and $H(z)\sigma_s$ are the power
spectrum and the amplitude of initial velocity perturbations,
$\Omega_m$ and $\Omega_\Lambda$ are matter and Dark Energy density
parameters. For both our simulations with the moderate resolution we
have approximately
\[
\sigma_s\approx {2.07\over\Omega_mh^2}\sigma_8{\rm Mpc},\quad
B\approx 1.27[1+(1+z)^3]^{-1/3}\,,
\]
and for $z\geq 1$ we get
\[
H(z)\approx H_0\sqrt{\Omega_m}(1+z)^{3/2},\quad B(z)\approx
1.27(1+z)^{-1}\,,
\]
\be
\sigma_U\approx 353 {\rm km/s}(1+z)^{-1/2}(\sigma_8/0.9)\,.
\label{U}
\ee

Using the matter velocities in the simulation $S_{150}$ we get for
the 1D velocity dispersion of clouds at $z\geq 1$
\be
\sigma_U=325km/s\,(1+z)^{-1/2}(1\pm 0.04)\,,
\label{Us}
\ee
and for the high resolution simulation $S_{50}$ we get
\be
\sigma_U=315km/s\,(1+z)^{-1/2}(1\pm 0.04)\,.
\label{u50}
\ee
The differences ($\sim 8\%\,\&\,11\%$) between the
expected (\ref{U}) and simulated (\ref{Us}, \ref{u50}) dispersions
characterize the impact of the large scale perturbations suppressed
in simulations owing to the finite box size.

For clouds selected in the simulation $S_{150}$ with the threshold
overdensity $\delta_{thr}=1.76$ and richness $3\cdot 10^{12}
hM_\odot\leq M_{cl}\leq 5\cdot 10^{14}hM_\odot$ at redshifts $z=1,
2\,\&\,3$ the probability distribution function (PDFs) for the cloud
velocities, $P_U(|U_j|/\sigma_U)$, $j=1, 2, 3$ are similar to each
other with high precision and are well fitted by the function \be
P_U(x_j)=0.77\exp(-x_j^2/2),\quad x_j=|U_j|/\sigma_U\,. \label{ufit}
\ee Small variations around the Gaussian fit (\ref{ufit}) depend on
peculiarities of the selected sample of clouds and can also
indicate the weak mass dependence of the measured dispersions.

\subsubsection{The matter fraction accumulated by the LSS}

To test the simulations and our clouds selection algorithm we
compare the expected and measured matter fraction accumulated by the
LSS. According to the Zel'dovich theory the fraction of matter
expanding along all three axes is $f_{exp}\approx0.5^3=0.125$ and,
therefore, the fraction of matter accumulated by all LSS elements is
\be f_{tot}\approx 0.875(1-z/z_f),\quad z\ll z_f\,. \label{texpec}
\ee This approximate relation slightly overestimates the total mass
accumulated by pancakes, filaments and clouds. More accurate but
multiparametric and more complex estimates can be found in
Demia\'nski \& Doroshkevich (1999, 2004).

For the simulation $S_{150}$ we get for $\delta_{thr}=1$
at $z\leq 3$
\be
f(M\geq 5\cdot 10^{11}h^{-1}M_\odot)\approx 0.73(1-z/5.7)\,.
\label{twofrac}
\ee
For the high resolution simulation $S_{50}$ we see earlier
matter concentration and for $\delta_{thr}=1$,
at $z\leq 6$, we get
\be
f(M\geq 2.5\cdot 10^{9}h^{-1}M_\odot)\approx 0.80(1-z/10)\,,
\label{frac50}
\ee
\[
f(M\geq 5\cdot 10^{11}h^{-1}M_\odot)\approx 0.67(1-z/10)\,.
\]
The moderate differences between the measured (\ref{twofrac},
\ref{frac50}) and expected (\ref{texpec}) values
demonstrate the moderate impact of the used selection parameters
used, mostly the minimal mass.
The earlier matter condensation in the high resolution
simulation $S_{50}$ (\ref{frac50}) as compared with the
simulation $S_{150}$ (\ref{twofrac}) illustrates the impact of
the small scale perturbations represented by the power spectrum
of simulation $S_{50}$.

\subsubsection{The mass function of clouds}

For high density virialized DM clouds the Press--Schechter mass
distribution is usually reproduced in simulations. However, we
will see below that for strongly anisotropic partly relaxed clouds
selected with moderate overdensity the simulated mass functions
are well approximated by a power law. Simple qualitative explanation
of such distribution is offered by the Zel'dovich theory.

According to this theory the process of the compact object
formation is determined by three conditions
\[
\Delta q_i-B(z)\Delta S_i = 0,\quad i=1, 2, 3\,,
\]
where the function $B(z)\leq 1$ was introduced in (\ref{sigu}) and
$\Delta S_i$ are the relative displacements of two points
separated by the comoving distance $\Delta q_i$ along $i^{th}$
principal direction. This means that the expected collapsed
volume, $V_q$, and corresponding mass, $M_{zt}$, can be
qualitatively estimated as
 \be
M_{zt}\propto V_q=\Delta q_1\Delta q_2\Delta q_3\propto V_s=\Delta
S_1\Delta S_2\Delta S_3\,,
\label{zm}
\ee
\[
\Delta S_i>q_{min}\geq 0\,.
\]
Here this condition assures that the collapse proceeds along all
three directions and restricts the volume $V_s$.

For any $M_0$
the probability to find the mass of collapsed clouds in the range
$M_0-\Delta M\leq M_{zt}\leq M_0+\Delta M$
is proportional to the probability to find $\Delta S_i\geq
q_{min}\geq 0$, and the volume $V_s$ in corresponding range
$V_0-\Delta V\leq V_s\leq V_0+\Delta V$. This simple approach
allows us to characterize the shape of the PDF $P(V_s)\propto
P(M_{zt})$ but its
normalization (in \ref{zm}) requires additional assumptions. More
detailed description of the complicated process of clouds formation
inevitably leads to much more complex multiparametric expressions
(see, e.g., Demia\'nski \& Doroshkevich 2004).

In turn, the PDF $P(V_s)$ can be directly obtained from the PDF
$P(\Delta S_i)$. For this purpose we use the simple three step numerical
method: 1. to construct the 3D sample of $\Delta S_i$ in accordance
with the theoretical PDF $P(\Delta S_i)$; 2. to select 1D subsample
of $V_s$ in accordance with the conditions (\ref{zm}); 3. to construct
the PDF $P(V_s)$.

The  PDF $P(\Delta S_i)$ is Gaussian \,
(Demia\'nski \& Doroshkevich 1999, 2004) with dispersion
$\sigma_S(\Delta q)$ and weak correlations,
\[
\xi_{ik}=\langle\Delta S_i\Delta S_k\rangle/\sigma_S^2\leq 1/3,
\quad i\neq k\,.
\]
With this PDF we constructed $10^6$ realizations of $\Delta S_i$
and found that the PDF $P(V_s)$ can be fitted by a power law $P(V_s)
\propto V_s^\kappa$. The slope $\kappa$ slightly varies depending
on the considered range of $V_s\geq V_{min}$. Thus, for
three progressively decreasing $V_{min}=V_1,\,0.1V_1,\,\&\, 0.01V_1$
the selected subsamples contain 30\,407, 50\,025 and 164\,103
elements with the slop of PDFs $\kappa\sim -2.5,\,-1.8\,\,\&\,-1.5$,
respectively. These variations indicate that this slop increases
for larger $V_s$.

In all simulations considered below the PDFs of masses of selected
clouds are fitted by power laws with $\kappa= - 2,\, - 2.4,\, - 2.2$
(\ref{mmfit}, \ref{Pp500}, \ref{pp50}) what is in the range of
expected values $-1.8 \geq \kappa \geq -2.5$. These variations of
$\kappa$ indicate mainly the impact of the resolution achieved in
simulations and the range of selected masses.

\section{Properties of simulated DM  LSS}

As is well known formation of the LSS  begins at high redshifts and
its evolution continues up to now. The matter fraction accumulated
within the LSS elements increases with time owing to both the
accretion of matter on the earlier formed LSS elements and creation
of new low mass LSS elements and bridges leading to the successive
integration of the LSS elements into multiconnected richer ones.
 At small redshifts these processes lead to the creation of the
joint network of compressed matter. The high efficiency of clouds
integration demonstrates the correlated character of the LSS
formation and the interaction of small and large scale perturbations.
Here we present some quantitative characteristics of these
processes obtained for the LSS formed by the DM particles only.

In some important aspects these clouds differ from
the LSS formed by both the DM and baryonic components and
observed as the Ly-$\alpha$ forest. In turn, both these
structures differ from the LSS formed by the luminous matter
(galaxies) and observed in large galactic surveys. None the
less, all these LSS elements are dominated by the DM component and
their evolution is driven by the same initial perturbations.
So, it can be expected that the most fundamental properties
of the LSS evolution can be established through the analysis
of the simulated DM LSS.

In this section we consider the clouds selected with the threshold
overdensity, $\delta_{thr}=1.76$, what allows us to partly suppress
formation of the very massive multiconnected clouds
at redshifts $z\leq 2$. Further on we will present results
obtained for the sample of compact LSS elements with DM masses
$10^{12}M_\odot\leq M\leq 5\cdot 10^{14}M_\odot$.

For the high resolution simulation $S_{50}$ we can obtain
more detailed description of the process of matter condensation
within various samples of high density clouds. Thus, we have
found that within the LSS clouds there is a population of
high density cores with $\delta\geq 10$. Such early
formation of high density low mass clouds was found also
by Diemand et al. (2008).

\subsection{Basic characteristics of compact clouds}

The set of selected compact clouds can be characterized by the
redshift evolution of their comoving number density, $\langle
n_{cls}(z)\rangle$, masses, $\langle M_{cl}(z) \rangle$, the
fraction of particles accumulated by clouds, $f_{com}$,  and
three comoving sizes $\langle L_i(z)\rangle$ ($i=1,2,3$)
determined along the three principal axes of each cloud.

For the sample of $N_{cls}\sim 11\,000 - 14\,000$ clouds
selected in the simulation $S_{150}$ with $\delta_{thr}=1.76$
and $M_{min}\leq M_{cl}\leq M_{max}$, we have with variations
smaller than10\%:
\[
\langle n_{cls}(z)\rangle\approx 3.6\cdot 10^{-3}{\rm Mpc^{-3}},
\quad \langle M_{cl}(z)\rangle\approx 3\cdot 10^{12}h^{-1}M_\odot\,,
\]

\[
f_{com}=\langle n_{cls}\rangle\langle M_{cl}\rangle/\langle
\rho\rangle\approx 0.13\,,
\]
\be
\langle L_i(z)\rangle\approx \ell_i\sqrt{\nu},\quad \nu=M_{cl}/
\langle M_{cl}(z)\rangle\,,
\label{lwh}
\ee
where $\ell_i$ are reduced cloud sizes with
\[
\langle\ell_1\rangle=1h^{-1}Mpc,\quad \langle\ell_2
\rangle=0.5h^{-1}Mpc,\quad \langle\ell_3\rangle=0.2h^{-1}Mpc\,.
\]
All these parameters characterize the selected sample of the LSS
elements. Mass dependence of the cloud sizes reflects the
complex structure of selected clouds and implies their partial
correlation. Their weak dependence on redshift characterizes the
balance between creation of new elements and matter outflow to
the more massive complex LSS elements caused by creations of
bridges between earlier formed clouds.

\begin{figure}
\centering
\epsfxsize=7.cm
\epsfbox{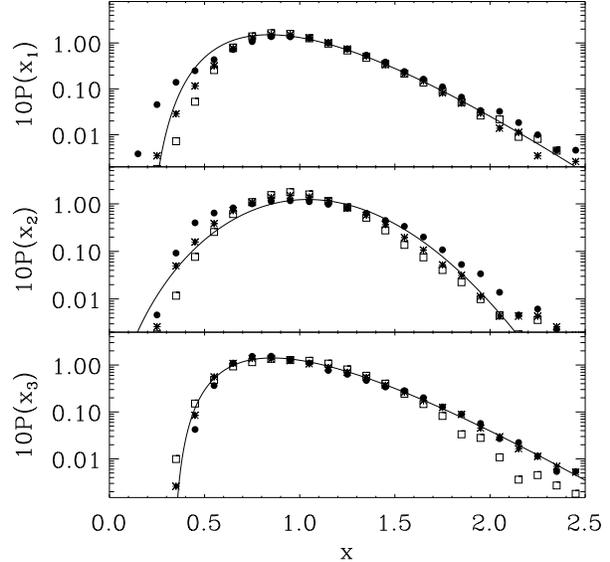}
\vspace{1.cm} \caption{The PDFs of clouds sizes $\ell_i$ along the
longest, middle, and shortest principal axes (top, middle and bottom
panels) are plotted for redshifts $z=1, 2\,\&\,3$ (points, stars and
squares) for clouds selected in simulation $S_{150}$ with the
threshold overdensity $\delta=1.76$ and richness given by
(\ref{nm}). Here $x_i=\ell_i/\langle\ell_i \rangle$. }
\label{flwh}
\end{figure}

The PDFs of the functions $\ell_i$ plotted in Fig. \ref{flwh} are
also weakly dependent upon redshift. The exponential decline of
the PDFs at $x_i=\ell_i/\langle\ell_i \rangle\geq 1$ reflects
the rare formation of asymmetric objects with large $\ell_i$
while at $x_i\leq 1$ the shape of the PDFs depends upon conditions
used for the cloud selection.

The PDF of cloud masses, $P_p(M_{cl}/\langle M_{cl}\rangle)$,
is surprisingly independent from redshift but
depends upon the range of considered velocity dispersion of
compressed matter (see Sec. 3.5). For all clouds it can be
approximated by the power law
\be
P_p(\nu)=0.25 \nu^{-2},\quad \nu=M_{cl}/\langle M_{cl}\rangle\,,
\label{mmfit}
\ee
what is close to predictions of the Zel'dovich theory (sec. 2.3.3).
It is interesting that here we do not see the exponential
decline of the PDF, $P_p(x)$, predicted by all theoretical models
and observed in the luminosity function of galaxies. In the
simulations such decline appears only at higher redshifts for
clouds selected with larger threshold overdensity. It can be
expected that these differences reflect the complex structure
of richer clouds formed partly by integration of earlier formed
clouds due to origin of low density bridges between them.

Let us note again that both the mean values $\langle L_i
\rangle$ and PDFs for the cloud sizes, $P(x_i=\ell_i/\langle
\ell_i\rangle)$, plotted in Fig. \ref{flwh} do not change
with redshift. These results
show that for so selected clouds, their mean sizes in real space
increase with time $\propto (1+z)^{-1}$ along each of the three
principal axes. This means that the mean cloud overdensity does
not change and, therefore, the possible dissipation of the LSS
elements owing to the matter expansion is not so essential.

The characteristics of compact clouds obtained in the simulation
$S_{500}$ are quite similar to those presented above. Let us only
note that for this simulation we find some excess of low mass clouds
as compared with the simulation $S_{150}$.
Indeed, instead of (\ref{mmfit}) we get for this simulation the PDF
\be
P_p(\nu)\propto \nu^{-2.4},\quad \nu=M_{cl}/\langle M_{cl}
\rangle\,,
\label{Pp500}
\ee
what confirms some excess of less massive clouds. This PDF differs
from (\ref{mmfit}) but it is in the range of predictions of the
Zel'dovich theory (sec. 2.3.3) and is quite
similar to that obtained below for colder less massive clouds
(\ref{cold}). This difference shows that some properties of the
selected clouds are sensitive to such factors as the possible
impact of the gaseous component, the initial realization of
perturbations, and the criteria of the clouds selection.

Analysis of the high resolution simulation $S_{50}$ allows
us to obtain more detailed description of the evolution of
compact clouds. As was noted above in this simulation we see
formation of the rich population of high density cores
with the overdensity $\delta_{thr}\geq 10$ and $M_{cl}\geq
10^{10}M_\odot$.
These cores have steep density profiles $\rho\propto r^{-2}$
and accumulate up to 40\% of matter at $z=1$. They are usually
incorporated into less dense clouds.

For the selected LSS elements we can trace their evolution
in a wider range of redshifts, $1\leq z\leq 6$. Thus, instead of
(\ref{lwh}) we get
\[
\langle n_{cls}(z)\rangle\approx 6.3\cdot 10^{-3}(1-0.1z)
{\rm Mpc^{-3}}\,,
\]
\[
\langle M(z) \rangle\approx 2.3\cdot 10^{12}(1-0.07z)h^{-1} M_\odot,
\]
\[
f_{com}\approx 0.18(1-0.16z)\,,
\]
\be
L_i=\ell_i\sqrt{\nu},\quad \nu=M_{cl}/\langle M_{cl}(z)\rangle\,,
\label{lwh50}
\ee
\[
\langle\ell_1\rangle=0.65(1+0.25z)h^{-1}{\rm Mpc}\,,
\]
\[
\langle
\ell_2\rangle=0.2(1+0.5z)h^{-1}{\rm Mpc}\,,
\]
\[\langle\ell_3\rangle=0.1(1+0.5z)h^{-1}{\rm Mpc}\,.
\]

At $z=3$ these sizes are quite similar to (\ref{lwh}) but at all
redshifts we see a slow continuous matter compression. This
difference reflects both the impact of small scale perturbations and
a more detailed description of the LSS evolution achieved in the
high resolution simulation.

The PDFs for the cluster sizes and their masses are
also similar to the previous ones. In particular,
\be
P_p(\nu)\propto \left\{
\begin{array}{ccc}
 \nu^{-2.0},& z \leq 3,&\cr
 \nu^{-2.2},& z>4\,,&\nu=M_{cl}/\langle M_{cl}\rangle\,,\cr
\end{array}\right.
\label{pp50}
\ee
what is also consistent with predictions of the Zel'dovich theory.
For larger threshold overdensity the theoretically expected
exponential decline of the mass function can be seen for high
$z$, when majority of clouds are well separated.
For the mass function of high density
halos selected within this simulation the exponential decline
is clearly seen at all redshifts.

\subsection{The velocity dispersion of compressed matter}

The velocity dispersion of particles within selected clouds depends
on three factors, which are: the retained Hubble expansion, the
enhanced random initial velocities of particles, and the velocity
created in the course of matter relaxation. All these factors are
closely linked with the clouds mass and therefore the measured
velocity dispersion also depends upon mass. The relative
contributions of these components vary with time but for
collisionless DM particles the rate of these variations depends upon
degree of matter compression and relaxation achieved along each of
the principal axes. This means that the velocity dispersions along
these axes are different and must be determined separately along the
longest, $\sigma_1$, middle, $\sigma_2$, and shortest, $\sigma_3$,
principal axes of each cloud.

\begin{figure}
\centering
\epsfxsize=7.cm
\epsfbox{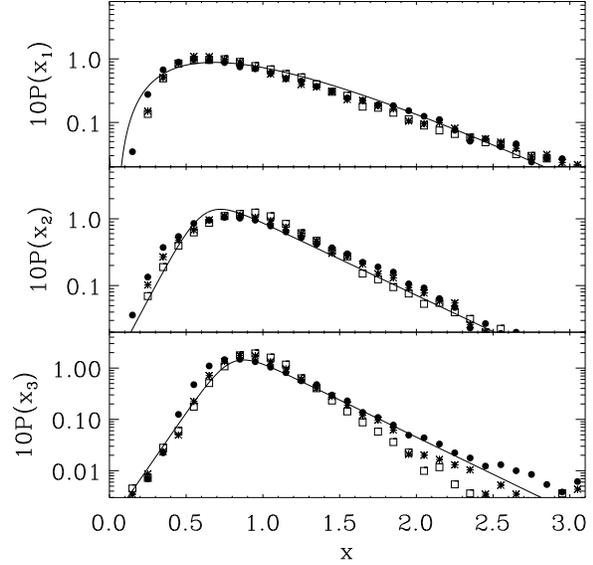}
\vspace{1.cm}
\caption{For the simulation
$S_{150}$ the PDFs of the reduced velocity dispersion, $c_{1,2,3}$,
along the longest, middle and shortest principal axes (top, middle
and bottom panels) are plotted for redshifts $z=1, 2\,\&\,3$
(points, stars and squares) for clouds selected with the threshold
overdensity $\delta= 1.76$ and richness given by (\ref{nm}). Here
$x_k= c_k/\langle c_k\rangle,\, k=1,2,3$ . } \label{fv}
\end{figure}
It is important that these velocity dispersions are much
smaller than the dispersions $\sigma_U$ introduced in (\ref{Us}).
This difference is caused by the strong correlation of the
particle velocities at the scales of clouds. The simple
analysis shows that the dispersion of initial velocities
within clouds decreases with the cloud mass, what stimulates
the formation of low mass high density clouds.

For clouds selected in the simulation $S_{150}$ with $\delta_{thr}
=1.76$ and $M_{min}\leq M_{cl}\leq M_{max}$ the velocity
dispersions along three principal axes, $\sigma_i$, slowly
vary with time but they depend upon the cloud richness,
\be
\sigma_i=c_i\nu^{\alpha_i},\quad \alpha_1=0.5,\quad \alpha_2=0.4,
\quad \alpha_3=0.3\,.
\label{fitv}
\ee
The mean values of the reduced velocity dispersions $c_i$ are
\be
\langle c_1\rangle=58km/s,\quad \langle c_2\rangle=66km/s,\quad
\langle c_3\rangle=80km/s\,,
\label{c150}
\ee
and their PDFs, $P(x_i=c_i/\langle c_i\rangle)$, are plotted in
Fig. \ref{fv} for three redshifts $z=1,~2, ~3$. The very weak
redshift evolution of these PDFs verifies the self similar
character of the simulated LSS evolution.

These PDFs demonstrate the expected exponential decline
$P(x_i)\propto\exp(-x_i/0.3)$ for $x_i\geq 1$. At the same time,
there are cold clouds with $c_i\leq\langle c_i\rangle$.

For the simulation $S_{500}$ the velocity dispersions, $\langle
c_i\rangle$ are $\sim$ 20\% smaller than (\ref{fitv}). It is
consistent with the excess of low mass clouds noted above
(\ref{Pp500}) and confirms the sensitivity of the characteristics of
the low mass clouds to the mass resolution, impact of the gaseous
component and realization of the initial velocity field. The PDFs of
the velocity dispersions are also similar to those plotted in Fig.
\ref{fv}.

For the simulation $S_{50}$, for the same mass range, at redshifts
$1\leq z\leq 5$ the mean velocity dispersions are described by
the relation (\ref{fitv}) with
\[
\alpha_1=0.4,\,\alpha_{2,3}=0.3
\]
\be
\langle c_1\rangle\approx 105(1-0.15z)km/s,
\label{sigv50}
\ee
\[
\langle c_{2}\rangle\approx 95km/s,\quad  \langle c_{3}\rangle
\approx 80 km/s\,.
\]
Their PDFs are quite similar to those plotted in Fig. \ref{fv}.

\begin{figure}
\centering
\epsfxsize=7.cm
\epsfbox{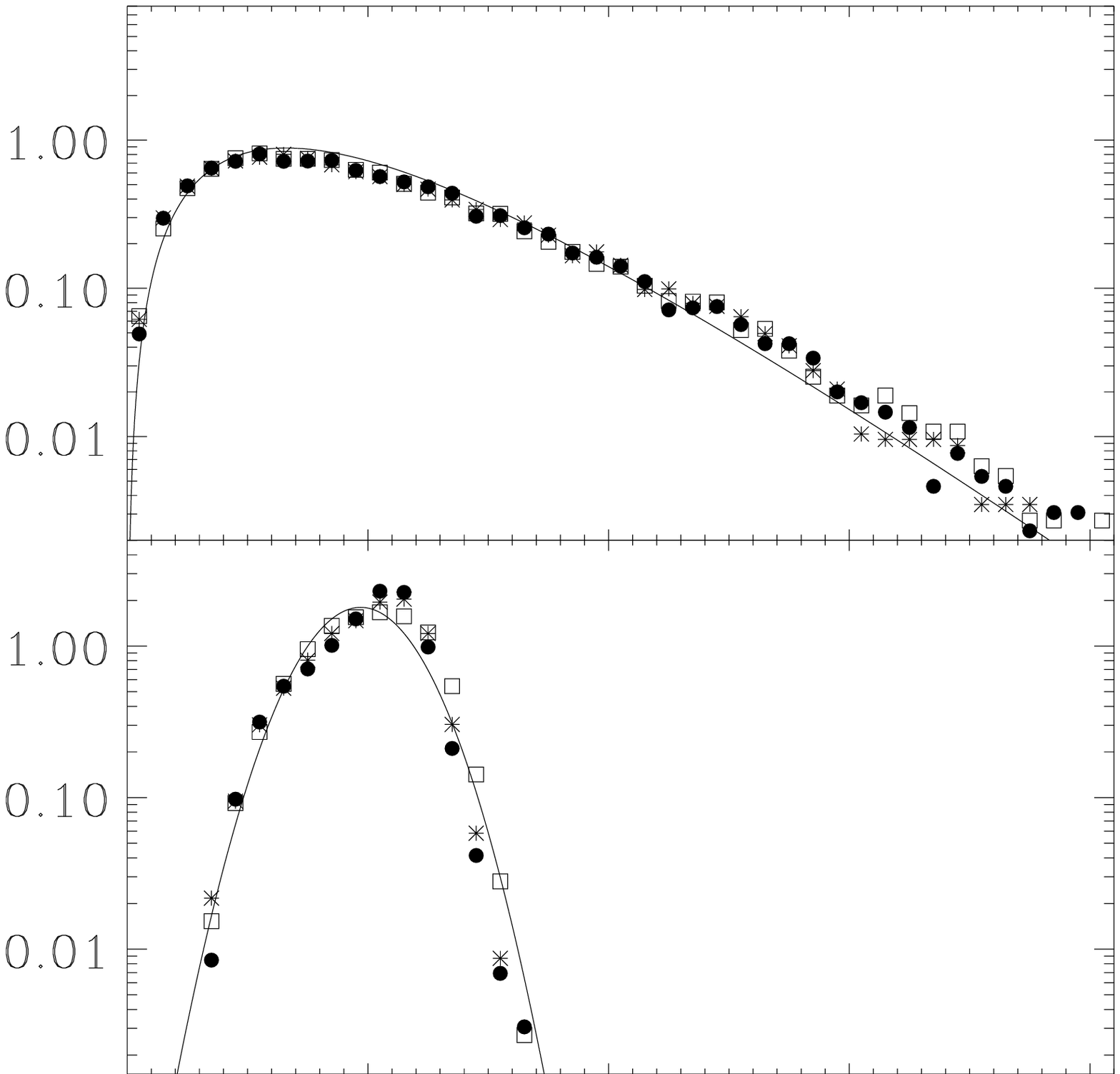}
\vspace{0.6cm} \caption{The PDFs of the reduced integral angular
momentum, $\mu$, and the reduced integral intensity of turbulent
motions, $\tau$, are plotted for redshifts $z=1, 2\,\&\,3$ (points,
stars and squares) for clusters selected with the threshold
overdensity $\delta= 1.76$ and richness given by (\ref{nm}). Here
$x_1=\mu/\langle\mu\rangle$, $x_2=\tau/\langle\tau\rangle$. }
\label{pgk}
\end{figure}

Differences between estimates (\ref{fitv}, \ref{c150}) and
(\ref{sigv50}) can be related to the influence of small scale
perturbations resulting in a more rapid relaxation and more
complex internal structure of selected compact clouds. This is
seen as the formation of high density cores within these clouds
and as a decrease of power indices $\alpha_1\,\&\,\alpha_2,_3$ in
(\ref{sigv50}) as compared with (\ref{fitv}). In
turn, action of these factors increases the reduced velocity
dispersions $c_1\,\&\,c_2$ (\ref{sigv50}) as compared with
(\ref{c150}). At the same time, for both simulations $\alpha_3$
and $c_3$ remain quite similar.

\subsection{Rotation and turbulent motions of the compressed
matter}

In the partly relaxed clouds particles move on closed orbits what
generates both the turbulent motions, $j_i$ (\ref{jturb}) and the
angular momenta of clouds as a whole, $J_i$ (\ref{turb}) along three
principal axes. Evidently, for isotropic simulations $\langle
J_i\rangle=0$ and further on we will consider characteristics of the
absolute values of angular momenta $|J_i|$ only. As before, these
functions depend upon cloud richness. For clouds selected in the
simulation $S_{150}$ we have
\be
|J_i|=\mu_ig_i\quad j_i=\tau_ig_i\,,
\label{fitj}
\ee
\[
g_1^2=(L_2^2+L_3^2)(\sigma_2^2+\sigma_3^2),\quad g_{2,3}^2=
(L_1^2+L_{3,2}^2)(\sigma_1^2+\sigma_{3,2}^2)\,,
\]
where $\mu_i\,\&\,\tau_i$ characterize dimensionless reduced angular
momenta and intensity of the turbulent motions, $L_i\,\&\,\sigma_i$
(\ref{lwh}\,\&\,\ref{fitv}) are the size and velocity dispersion of
clouds. The mean values of the reduced angular momenta are found to
be independent of $z$ and \be \langle \mu_1\rangle=0.08,\quad
\langle \mu_2\rangle=0.15, \quad \langle \mu_3\rangle=0.15\,.
\label{jmns}
 \ee
The small clouds rotation around the longest axis and exponential
shapes of PDFs \be P(x_i)\approx 0.12\exp(-x_i),\quad
x_i=\mu_i/\langle\mu_i \rangle\,, \label{jpdf} \ee are consistent
with the TTT predictions.

The mean values of the reduced intensity of turbulent motions are
also found to be independent of $z$ and \be
\langle\tau_1\rangle=0.64,\quad \langle\tau_2\rangle=0.76,\quad
\langle\tau_3\rangle=0.65\,, \label{tmns} \ee and their PDFs are
fitted by the functions \be P(x_i)\approx
0.14\exp[-(x_i-1)^2/0.14],\quad x_i=\tau_i/ \langle\tau_i\rangle\,.
\label{tpdf} \ee

For the total momentum of the clouds, $J$, and for the
total intensity of turbulent motions, $j$,
\be
J=\sqrt{J_1^2+J_2^2+J_3^2}=\mu g_t,\quad j=\sqrt{j_1^2+j_2^2
+j_3^2}=\tau g_t\,,
\label{jj}
\ee
\[
g^2_t=(L_1^2+L_2^2+L_3^2)(\sigma_1^2+\sigma_2^2+\sigma_3^2)\,,
\]
we get \be \langle\mu\rangle=0.17,\quad \langle\tau\rangle=0.8\,.
\label{imns} \ee The PDFs for the functions $\mu\,\&\,\tau$ are
plotted in Fig. \ref{pgk}. The obtained value of $\langle
\tau\rangle$ corresponds to domination of  elliptical trajectories
of particles with the ratio of axes $b/a \approx 0.57$, what is
consistent with the ratios of the cloud sizes (\ref{lwh}).

For the simulation $S_{50}$ we have, at $1\leq z\leq 6$
and for the same mass range,
instead of (\ref{jmns}), (\ref{tmns}) and (\ref{imns}):
\[
\langle \mu_1\rangle=0.08(1-0.07z),\quad
\langle\tau_1\rangle=0.57,
\]
\be
\langle \mu_2\rangle=0.1(1-0.07z),\quad
\langle\tau_2\rangle=0.60(1-0.08z),
\label{tm50}
\ee
\[
\langle \mu_3\rangle=0.09(1-0.1z),\quad
\langle\tau_3\rangle=0.57(1-0.09z)\,,
\]
\[
\langle\mu\rangle=0.13(1-0.08z),
\quad \langle\tau\rangle=0.7(1-0.08z)\,.
\]

Comparison of $\langle\mu\rangle$ with $\langle\tau\rangle$ in
(\ref{jmns}), (\ref{tmns}), (\ref{imns}) and (\ref{tm50})
shows that only $\sim 15 - 20\%$ of the angular momenta of
individual particles are transformed into the angular momentum
of a cloud as a whole, what characterizes the efficiency of the
generation of angular momentum according to the TTT.

These results are also consistent with predictions of the
Zel'dovich theory (\ref{Jth} -- \ref{zturb}) and indicate
the significant
contribution of initial random velocities in the turbulent
motions within compressed clouds.

\subsection{Relaxation of the compressed matter}

\begin{figure}
\centering
\epsfxsize=7.cm
\epsfbox{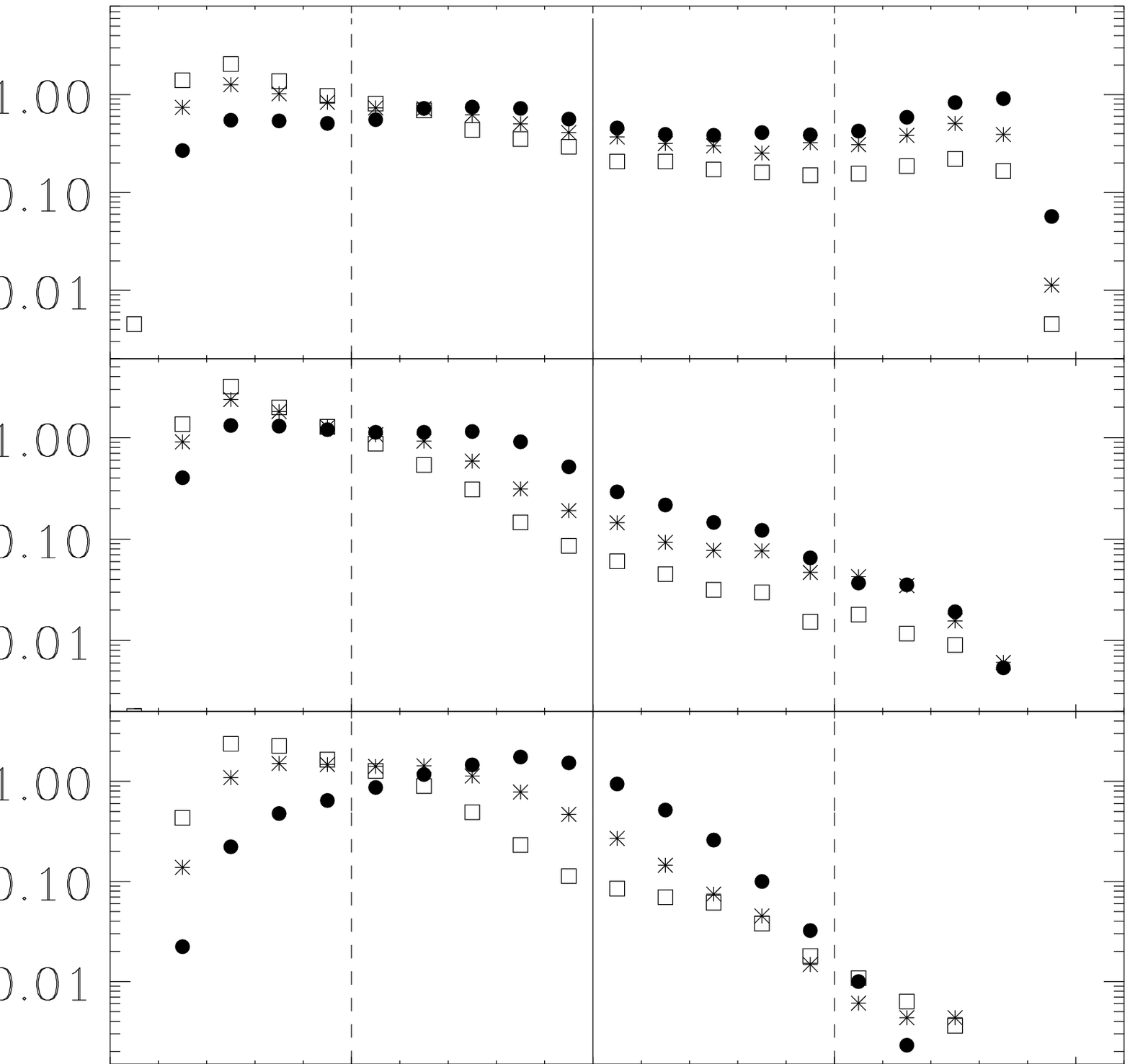}
\vspace{1.cm} \caption{For the simulation
$S_{150}$ the PDFs of the ratio of projected velocity and velocity
dispersion along the longest, $w_1$, middle, $w_2$, and shortest
$w_3$, axes (top, middle and bottom panels) are plotted for
redshifts $z=1, 2\, \&\,3$ (points, stars and squares) for clouds
selected with the threshold overdensity $\delta=1.76$ and richness
given by (\ref{nm})}.
\label{fw}
\end{figure}

The three parameters $w_1, w_2\,\&\,w_3$ introduced in
(\ref{drelax}) allow us to discriminate between the regular
expansion and compression of matter along the principal axes and to
estimate the degree of matter relaxation.  As noted in Sec. 2.2,
$w_i\geq 1$ indicates domination of matter expansion, while  $w_i\leq
1$ corresponds to domination of matter compression along the $i^{th}$
principal axis. For the simulation $S_{150}$ and for redshifts $z=1,
2, 3$ the mean values $\langle w_i\rangle$ are listed in Table 1 and
their PDFs are plotted in Fig. \ref{fw}. These results indicate that
for our samples of selected clouds the matter compression along all
three axes dominates but it is progressively decelerated with time.
The symmetry of the PDF $P(w_1)$ points out that for our selected
clouds both the compression and expansion along the longest axis are
almost equally typical. These estimates only weakly depend upon the
richness of clouds.

Another manifestation of the retained expansion along the
longest axis is the tight correlation of velocity dispersion
and size:
\[
 \sigma_1\propto L_1^{0.8},
\]
with the deviation of $\sim$10\%. This expression is
quite similar to the Hubble flow. For the middle and
the shortest axes such correlations are very weak.

These results confirm that along the longest principal axis the
relatively slow relaxation is accompanied by the retained expansion
of clouds, which persists for some time after cloud formation. In
contrast, along the middle and shortest principal axes the moderate
retained compression of clouds predominates. These results are
consistent with the strongest matter compression along the shortest
principal axis (which is responsible for pancakes formation) and the
successive transformation of prolate pancake -- like clouds into
elongated filamentary -- like ones.

\begin{table}
\caption{Degree of relaxation of selected clouds}
\label{tbl1}
\begin{tabular}{llll}
\hline
        &z=1&z=2&z=3\cr
\hline
$\langle w_1\rangle$-1&0.05&-0.2&-0.42\cr
$\langle w_2\rangle$-1&-0.4&-0.53&-0.62\cr
$\langle w_3\rangle$-1&-0.2&-0.42&-0.55\cr
$f_{rel}$              &0.8    &0.44 &0.2\cr
\hline
\end{tabular}
\end{table}

To estimate the degree of matter relaxation along the shorter axis
we can use the fraction of clouds, $f_{rel}(z)$, for which
$|1-w_3(z)|\leq 0.5$. For three redshifts the estimates of this
fraction are listed in Table 1. The rapid growth of this fraction
with time illustrates the progressive relaxation of the compressed
matter along the shortest principal axis. The fraction of matter
relaxed along the longest principal axis is only half of that along
the other two axes. For the simulation $S_{500}$ the functions
$\langle w_i(z)\rangle - 1$ and $f_r(z)$ are similar to the ones
discussed above.

 More complex internal structure of clouds selected from the
high resolution simulation $S_{50}$ leads to a more rapid
relaxation of the compressed matter. Thus, we have for the
matter fraction with  $|1-w_3|\leq 0.5$
\be
f_{rel}(z=5)\approx 0.6,\quad f_{rel}(z=3)\rightarrow 0.9\,.
\label{relax-50}
\ee
These results illustrate the impact of small scale perturbations
and strong links between the resolution achieved in the simulation
and the processes of clouds formation and relaxation.

\subsection{Colder and hotter clouds}

The strong link between the velocity dispersions and richness
of clouds (\ref{fitv}) illustrates the large scale correlation
of the initial velocity field. None the less, the scatter around
the mean dispersions (\ref{fitv}) is large and differences
between properties of colder and hotter clouds deserve
special investigation.

In the simulation $S_{150}$ the subpopulation of colder clouds was
selected from the full sample of clouds by the condition that
$\sigma_3 \leq 40km/s$. It contains about half of clouds with the
lower richness and the steeper mass function $P_c(M_{cl}/\langle
M_{cl} \rangle)$: \be \langle M_{cl}(z)\rangle\approx 7\cdot
10^{11}h^{-1}M_\odot,\quad P_c(M_{cl})\propto [M_{cl}/\langle
M_{cl}\rangle]^{-2.7}\,. \label{cold} \ee

The subpopulation of hot rich clouds with $\sigma_3\geq 40km/s$,
is characterized by larger mean richness  and
shallower mass function, $P_h(M_{cl}(z)/\langle M_{cl}(z)\rangle)$,
\be
\langle M_{cl}(z)\rangle\approx 3.2\cdot 10^{12}h^{-1}M_\odot,\quad
P_h(M_{cl})\propto [M_{cl}/\langle M_{cl}\rangle]^{-1.8}\,.
\label{rich}
\ee

Both colder and hotter subpopulations are dominated by partly
relaxed clouds expanded along the longest principal axis and
compressed along the middle and shortest axes and for them the
numerical estimates of $\langle w_i\rangle$ are quite similar to the
ones listed in Table 1. For the subpopulations of hotter clouds the
fraction of relaxed clouds with $|1-w_3(z)|\leq 0.5$ is close to
$f_{rel}$ listed in Table 1, while for the colder clouds this
fraction is smaller by a half.

\section{Core-sampling approach.}

For a more detailed comparison of the simulated matter distribution
with the observed characteristics of the LSS and the Ly--$\alpha$
forest it is convenient to use the core-sampling approach
(Doroshkevich et al. 2001, 2004; Demia\'nski et al. 2006). With this
approach we retain only particles within the selected sample of
clouds, and divide the simulated box into a system of rectangular
cores of comoving size $d_{core}\times d_{core}\times L_{box}$, and
consider properties of particles situated within each core.

All particles within a core are projected on the core axis and
using the 1D cluster analysis are collected into clumps.
For so selected clumps we can determine several
characteristics such as the mass, $M_{cr}$, and the comoving
surface density, $q_{cr}=M_{cr}/d_{core}^2$. Other characteristics
relate to the clump position and velocity along the core.
They are the comoving clump separation, $D_{sep}$, and their size
(thickness), $t_{cr}$, and overdensity, $\delta_{cr}=q_{cr}/t_{cr}
/\langle\rho\rangle$, the relative velocity of neighbouring clumps,
$|\delta U|$, the dispersions of clump velocity, $\sigma_U$, and
particle velocity within clumps, $\sigma_v$. We can also estimate
the degree of matter relaxation, $w_{cr}$, (\ref{drelax}), and
the dispersion of transverse velocity of matter within clumps,
$\sigma_{tr}$.

The core--sampling analysis utilizes four parameters, namely, the
comoving size of the core, $d_{core}$, the comoving threshold
linking length used in the 1D cluster analysis, $l_{thr}$, and the
minimal and maximal richness of clumps, $q_{thr}\,\&\, q_{max}$,
retained for the analysis. Results of the analysis depend mainly
upon the choice of $d_{core}$. Two last parameters allow us to
remove from the sample poor clumps that arise owing to random
intersection of the core and a cloud periphery, and small number of
extremely rich clumps with $q\geq q_{max}$, what strongly distorts
the mean measured characteristics.

Here we apply the
core--sampling approach to investigate the full sample of clouds
selected in the simulation $S_{150}$  with the threshold overdensity
$\delta_{thr}=1.76$. In our analysis, we use
\be
d_{core}=1.h^{-1}Mpc,\quad l_{thr}=0.5h^{-1}Mpc\,,
\label{1dthr}
\ee
\[
q_{thr}=M_{cs}/d_{core}^2=3\cdot 10^{11}M_\odot hMpc^{-2},\quad
q_{max}=10q_{thr}\,.
\]
These samples contain $\sim (1.3 - 1.5)\cdot 10^5$ clumps.
Naturally, variations of these parameters change the fraction of low
and high mass clumps in the selected sample and also the main
characteristics of the sample of clumps. However for a wide set of
limits (\ref{1dthr}) we see a weak redshift dependence of the mean
characteristics and their PDFs, what confirms the self similar
character of the LSS evolution.

In many respect this approach and its limitations are similar to
those appearing in the analysis of the Ly--$\alpha$ forest
(Demia\'nski et al. 2006) and the pencil beam observations of the
LSS (Doroshkevich et al. 2001, 2004). Of course the criteria of
clump selection in observations and simulations are not identical.
Indeed, if the Doppler parameter of the forest line coincides with
the velocity dispersion $\sigma_v$, then the evolutions of the DM
surface density $q(z)$  and the column density of neutral hydrogen,
$N_{HI}$, are very different as they are driven by different factors.
Thus, we cannot directly compare simulated results with observations.
However even such analysis allows us to reveal qualitative
similarity of the observed and simulated LSS evolution and to
demonstrate the impact of the main factors that influence evolution
of the observed LSS.

\begin{figure}
\centering
\epsfxsize=7.cm
\epsfbox{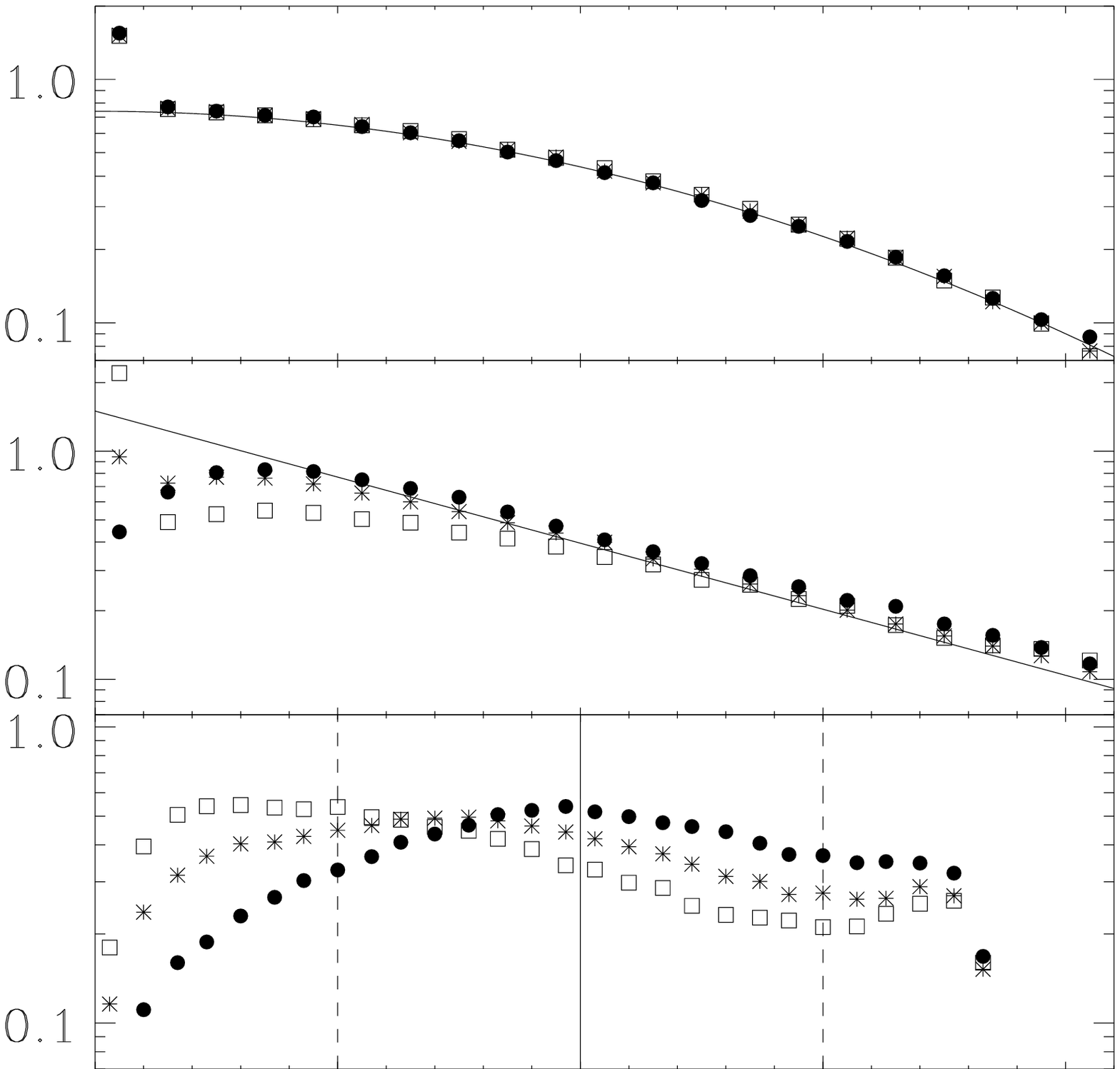}
\vspace{1.cm}
\caption{The PDFs of clusters
velocity, $P(x_u)$, the velocity dispersion within clusters,
$P(x_v)$, and the degree of relaxation, $P(w_{cr})$, are plotted for
redshifts $z=1, 2\,\&\,3$ (points, stars and squares) for clusters
selected within cores under conditions (\ref{1dthr}). Here $x_u=|U|/
\sigma_U$, $x_v=\sigma_v/\langle\sigma_v\rangle$.}
\label{uvw}
\end{figure}

\subsection{Dynamical characteristics of the DM LSS}

The redshift evolution of dynamical characteristics of the simulated
DM clumps obtained with the core sampling approach are found to be
similar to those obtained in the previous section. Thus, with the
threshold parameters (\ref{1dthr}) and for $3\geq z\geq 1$ we get
for the velocity dispersion of clumps, $\sigma_U$, and for the
relative velocity of neighboring clumps, $\delta U$ \be
\langle\sigma_U\rangle\approx 308km/s(1+z)^{-1/2}(1.\pm 0.05)\,,
\label{1Dmns} \ee \be \langle|\delta U|\rangle\approx
272km/s(1+z)^{-1/2}(1.\pm 0.03)\,. \label{delt_u} \ee So defined
$\sigma_U$ and the value (\ref{Us}) measured for all particles are
close to each other. The PDF for the dispersion of the 1D clumps
velocity, $P(x_u=|U|/\sigma_U)$, plotted in Fig. \ref{uvw} (top
panel) is close to the Gaussian one. However at all redshifts there
is some excess of clumps ($\sim 6\%$) with extremely low $\sigma_U$.

Both velocity dispersions of the compressed matter measured with
the core sampling approach along the core, $\langle\sigma_v\rangle$,
and in the transverse directions, $\langle\sigma_{tr}\rangle$,
are weakly dependent upon redshift, what is similar to the behavior
of the $\sigma_i$ (\ref{fitv}). Thus we get
\be
\langle\sigma_v\rangle\approx\langle\sigma_{tr}\rangle
\approx 49km/s(1\pm 0.2)\,.
\label{sig-tr}
\ee
The random orientation of the cloud principal axes and the probing
core leads to intermixture of velocities along three principal axes,
and the velocity dispersion measured in transverse directions,
$\langle\sigma_{tr}\rangle$, is the same as $\langle\sigma_v\rangle$
measured along the core.

The PDFs of the velocity dispersion of matter compressed
within clumps,  $P(x_v=\sigma_v/\langle\sigma_v\rangle)$, is
plotted in Fig. \ref{uvw} (middle panel) for three redshifts,
$z=1, 2, 3$. As is seen from this Figure for $x_v\geq 0.5$
these PDFs are quite similar to each other and they are well
fitted by the exponential function
\be
P(x_v)\approx 1.5 exp(-x_v/0.75)\,.
\label{fit_v}
\ee
However fraction of clumps with $x_v\leq 0.1$
progressively decreases with time from $\sim 25\%$
at z=3 and down to $10\%$ at redshifts $z=1$. This fraction can
be related to the low mass unrelaxed clumps formed due to the
intersection of cores with complex low density periphery of the
selected 3D clusters.

As before, the mean value $\langle w_{cr}\rangle\approx 1$
characterizes the symmetry of the matter compression and expansion
along the core in the sample of selected clumps. However, the PDF
$P(w_{cr})$ plotted in Fig. \ref{uvw} (bottom panel) characterizing
the degree of matter relaxation along the core shows some variation
with redshift. Using this PDF we can estimate the fraction of
clumps, $f_{rel}(z)$ with $0.5\leq w_{cr}\leq 1.5$  which are
almost relaxed along the core at redshifts $z=3, 2,\,\&\,1$ as
follows:
\be
f_{rel}(3)=0.56,\quad f_{rel}(2)=0.64,\quad f_{rel}(1)=0.7\,.
\label{1Drel}
\ee
These fractions are similar to the
fraction of clusters relaxed along the shortest axis listed in Table
1, what demonstrates moderate impact of the macroscopic motions along
the core. This result seems to be natural because of the moderate
difference between the fraction of matter relaxed along the shortest
and longest principal axes (Table \ref{tbl1}\,\&
\,Fig. \ref{fw}).

\subsection{Spatial characteristics of the DM LSS}

The simulated overdensity of clumps, $\langle\delta_{cr} \rangle$,
the surface density, $\langle q_{cr}(1+z)^{-2} \rangle$, and the
comoving thickness of clusters, $\langle t_{cr}\rangle$, are found
to be weakly dependent upon the redshift,
\[
\langle q_{cr}(1+z)^{-2}\rangle\approx 4\cdot 10^{11}\frac{hM_\odot}
{Mpc^2}(1.\pm 0.2)\,,
\]
\be
\langle t_{cr}\rangle\approx 1~h^{-1}{\rm Mpc}(1.\pm 0.1)\,,
\label{qt1}
\ee
\[
\langle\delta_{cr}\rangle=\left\langle{q_{cr}\over t_{cr}
\langle \rho\rangle(1+z)^2}\right\rangle\approx 5(1.\pm 0.15)\,,
\]
where $\langle\rho\rangle$ is the mean comoving
density (\ref{dthr}).

The redshift evolution of the comoving cluster separation,
\be
\langle D_{sep}\rangle\approx 16.2h^{-1}{\rm Mpc}(1.\pm 0.1)\,,
\label{dsep}
\ee
is also weakly dependent upon redshift but it strongly
depends upon threshold parameters (\ref{1dthr}) determining the
clumps selection procedure.

For the simulation $S_{50}$ the mean comoving separation between
clumps is
\be
 \langle D_{sep}\rangle\approx 7h^{-1}{\rm Mpc}\,.
 \label{dsep50}
\ee
The difference between this value and (\ref{dsep}) illustrates
the impact of the box size (the limited core length). The same
result as (\ref{dsep50}) is obtained when we cut a ``box'' of
50$h^{-1}{\rm Mpc}$ from the simulation $S_{150}$.

The PDFs of the DM surface density, $P(q_{cr}/\langle
q_{cr}\rangle)$, and of the objects separation, $P(D_{sep}/\langle
D_{sep}\rangle)$, also weakly depend on redshift  and are well
fitted by exponential functions
\be
P(x_q)\approx 0.8\exp(-x_q/0.8),\quad x_q=q_{cr}/\langle
q_{cr}\rangle\geq 0.2\,,
\label{fit-msep}
\ee
\[
P(x_{sep})\approx 0.9\exp(-x_{sep}),\quad x_{sep}=
D_{sep}/\langle D_{sep}\rangle\,.
\]

These results show that with the core--sampling approach the
redshift evolution of the simulated LSS is well characterized by the
slow regular variations of its mean characteristics, while the
corresponding PDFs practically do not change at all. This fact
verifies again the self similar character of the LSS evolution, at
least for its richer elements formed by the DM component of the
Universe.  It is consistent with the observed evolution of the
Ly-$\alpha$ forest (Demia\'nski et al. 2006).

\section{Summary and discussion}

Many branches and strongly nonhomogeneous matter distribution
typical for the richer LSS elements determine their complex
multiconnected structure and do not allow to characterize them in
any simple way. Thus, we are compelled to restrict our analysis
to the sample of compact DM clouds with a moderate richness and
overdensity. The simple procedure of the sample selection and the
basic characteristics of clouds are described in Sec. 2.  More
refined technique proposed for selection and discrimination of
the DM filaments and walls (see, e.g., discussion in Doroshkevich
et al. 2004, Aragon--Calvo et al. 2007; Zhang et al. 2009) can
be efficient mainly for the solution of some special problems.

In many respects evolution of the selected population
is typical for all LSS elements but some discussed
characteristics depend upon the procedure of sample selection.
First of all this relates to the spatial characteristics of
clouds.

As is well known the LSS evolution is determined by the accretion of
diffuse matter, creation of poorer clouds and integration of clouds
into larger ones. These processes lead to rapid matter concentration
within population of large multiconnected clouds and to successive
network formation. Thus, in the simulation $S_{150}$ the matter
fraction accumulated by richer clouds with $M_{cl}\geq
10^{14}h^{-1}M_\odot$ increases by a factor of 5 between redshifts
$z=3$ and $z=0$, while the fraction of particles accumulated by
clouds with $5\cdot 10^{11}h^{-1}M_\odot\leq N_p \leq 10^{14}
h^{-1}M_\odot$ increases by a factor of 1.5 only. This means that at
$z\leq 3$ the LSS evolution is dominated by successive integration
of the earlier formed LSS elements into the richest ones. The
evolution of compact clouds with moderate richness is determined by
a balanced action of coalescence and formation of new clouds.

The existence of the population of DM clouds with moderate
richness is in itself quite interesting. In many respects its
slow evolution leads to the stability of measured clouds
characteristics, which are distorted mainly by the successive
formation of high density cores. However, at higher redshifts
the influence of the process of integration of clouds becomes
less significant, the creation of new clouds and accretion
of diffuse matter dominate and characteristics of clouds
(\ref{lwh50}) evolve more rapidly.

The performed analysis of the simulated DM LSS allows us to clarify
some important factors that cause its evolution, and to link them
with the power spectrum of the initial perturbations and with the
observed evolution of the Ly--$\alpha$ forest. The limited mass
resolution  -- $M\sim 2\cdot 10^{10}M_\odot$ and $M\sim 7\cdot
10^{7}M_\odot$ in analyzed simulations -- prevents the formation of
low mass clouds comparable with the majority of the observed
Ly-$\alpha$ absorbers.
This means that our results cannot be directly compared with the
observed characteristics of the Ly--$\alpha$ forest. However, such
analysis allows us to find some characteristics of evolution of the
simulated DM LSS elements, what is very interesting in itself. It is
important also that the main basic features of this evolution are
similar in many respects to the observed properties of the
Ly--$\alpha$ forest, what allows us to explain some peculiarities of
the forest evolution discussed in the Introduction.

\subsection{Main inferences}

The most important features of the evolution of the selected
DM LSS elements can be summarized as follows:
\begin{enumerate}
\item{} Weak redshift variations of the basic characteristics
        of the selected sample of DM clouds such as their
        mean comoving size and velocity dispersion.
\item{} Measured PDFs of the basic characteristics of the
        LSS elements are weakly dependent upon the
        redshift, what implies the self similar character
        of the LSS evolution.
\item{} Significant degree of relaxation of the compressed matter
        along the shortest principal axis and retained matter
        expansion along the longest principal axis.
\item{} Weaker rotation of clouds along their longest axis.
\item{} Some measured clouds characteristics depend on the
procedure of clouds selection.
\end{enumerate}

\subsection{Impact of simulation parameters}

Comparison of characteristics of clouds obtained for three simulations
with different box sizes and resolutions shows that some of them are
sensitive to the resolution, the box size and the impact of baryonic
component. In particular, complex internal structure of clouds
selected in the high resolution simulation $S_{50}$ leads to more
complex evolution of the basic characteristics of clouds as compared
with the results obtained for lower resolution simulations. In the
simulation $S_{50}$ clouds are partially fragmented into high density
cores, what distorts their dynamical properties. The fraction of
high-density virialized clumps increases with resolution achieved
in simulation. However, even a large mass fraction (up to 40\%)
of these high-density clumps does not affect strongly the main
qualitative inferences: the self-similar evolution of
characteristics of clouds and exponential decline of their PDFs.

The most prominent impact of resolution is detected on the clouds
size along the middle and the shortest axes $\ell_{2,3}$
(\ref{lwh50}), the velocity dispersion along the longest principal
axis (\ref{sigv50}), and the relaxation parameter, while other
properties of clouds are quite similar in all three simulations.

In the simulation $S_{50}$ the box size affects separation between
clouds (\ref{dsep},\, \ref{dsep50}). The box size of 150 $h^{-1}$
Mpc should be enough for the stable reproduction of both these
properties.

\subsection{Self similarity of the DM LSS evolution}

The most interesting result of our analysis is the self similar
character of evolution of the DM LSS which is manifested at $z
\leq 4 -5$ as the surprising stability of PDFs of the basic
characteristics of clouds found with both the 3D analysis in
Sec. 3 and with the core--sampling approach in Sec. 4. This
result is obtained for three simulations performed in different
way and with different number of particles, resolution and
realization of the
initial particle positions and velocities. These features are
similar also to the redshift independence of PDFs of the
observed characteristics of the Ly--$\alpha$ forest
(Demia\'nski et al. 2006). These analogies point to the
universal character of the detected self similarity.

Other manifestation of this self similarity is the universality of
the NFW density profiles within simulated DM halos (see, e.g.,
Navarro, Frenk, White 1995, 1996, 1997). The self similar character
of the DM halos formation is seen as the regular redshift variations
of the internal structure of individual halos. However, the complex
process of formation and relaxation of the high density LSS elements
and halos can be described analytically as a self similar process
only for the simplest cases (see, e.g., Fillmore \& Goldreich, 1984;
Gurevich, \& Zybin, 1995; Sikivie, Tkachev, Wang Yun, 1997; Nusser,
2001).

For objects formed in the course of mildly nonlinear
compression the self similar character of evolution is
predicted by the Zel'dovich approximation (Zel'dovich 1970;
Shandarin \& Zel'dovich 1989), where the evolution is described
by a product of time and space dependent functions. The
statistical description of the DM LSS based on the Zel'dovich
theory (Demia\'nski \& Doroshkevich 1999, 2004) clearly
demonstrates the expected self similarity.

The simulated evolution of DM clouds is a strongly deterministic
process driven mostly by the  used realization of initial
perturbations. The statistical properties of clouds are determined
by the correlation and structure functions expressed through the
simulated power spectrum. Therefore we can expect that the self
similarity of the simulated evolution is caused by the properties of
the power spectrum responsible, in particular, for the formation of
clouds.

Such self similarity naturally appears for the power spectrum
approximated by a power law. Indeed, in this case all correlation
and structure functions are described by the time dependent
amplitudes and standard space dependent functions. This usually
leads to the self similar evolution of the basic statistical
characteristics of the DM LSS. The power spectra used in considered
here simulations are well approximated by a power law, what explains
the found self similarity.

Of course, this inference is related to the limited range of scales
of the power spectrum and therefore to the limited range of clouds
richness. However, we can expect that this range extends down to
the scales typical for the majority of the forest elements, what in turn
explains weak redshift variations of the observed PDFs of forest
characteristics. In contrast, the more complex process of formation of the largest
clouds and the network of the high density LSS elements is driven
by the large scale part of the power spectrum that determines the
formation of low density bridges between high density clouds.
It is convenient to analyze these mildly nonlinear processes on the
basis of the Zel'dovich approximation (Demia\'nski \& Doroshkevich
1999, 2004).

\subsection{Evolution of the velocity dispersions and the
degree of relaxation of clouds}

Second important result of our analysis relates to the
evolution of the internal structure of clouds. Both the
moderate mass resolution achieved in simulations and
moderate richness of the dominant population of clouds prevents
more detailed description of the anisotropic matter distribution
within clouds. However, high force resolution achieved in
the considered here simulations allows us to estimate, with a reasonable
precision, the evolution of the velocity dispersion and
the degree of relaxation of the compressed matter.

The mean values of the velocity dispersions $\sigma_1, \sigma_2
\,\&\,\sigma_3$, (\ref{fitv}) and even their PDFs presented in
Fig. \ref{fv} for three redshifts are quite similar to each
other. However, the PDFs of the degree of relaxation presented
in Fig. \ref{fw} demonstrate significant differences in
characteristics of the velocity components. Indeed, at all
redshifts the excess of larger values of $w_1\geq 1$ indicates
noticeable contribution of the retained matter expansion and moderate
degree of relaxation along the longest axis. At the same time,
the excess of smaller values of $w_2\leq 1$ demonstrates
domination of matter compression along the middle axis and
progressive transformation of a pancake like cloud into a filamentary
like one. The degree of relaxation along this axis is moderate
but it is larger than that along the longest axis.

In contrast, the more symmetric shape of the PDF $P(w_3)$ in
Fig. \ref{fw} indicates that similar fraction of matter
is expanded and compressed along the shortest axis. This
Figure shows progressive point concentration near the center,
what reflects the progressive randomization of velocities and
growth of the degree of relaxation of the compressed matter.
Evaluation of the fraction of relaxed clouds and the
corresponding matter fraction $f_{rel}$ (Table 1) characterizes
these processes quantitatively.

Similar characteristics of the velocity dispersions are also
obtained with the simulations  $S_{500}$ and $S_{50}$. The
most serious quantitative divergences are found for
characteristics of low mass clouds and are discussed in Sec. 3.

 It is interesting, that the estimated state of relaxation of
clouds with the core sampling approach in (\ref{1Drel}) is very
close to that obtained with the 3D analysis. This fact shows that
the random orientation of the clouds with respect to the line of
sight results in the intermixture of thermal and large scale bulk
velocities but does not distort strongly the measured velocity
dispersions and the degree of relaxation of compressed matter.

Special analysis confirms that the subpopulations of colder
and hotter clouds with $\sigma_3\leq\langle\sigma_3\rangle$
and $\sigma_3\geq\langle\sigma_3\rangle$ respectively,
contain comparable fractions of matter with different mean
richness and mass functions. Both subpopulations are dominated
by partly relaxed clouds slowly expanding along the longest
principal axis.
The colder clouds represent population of anisotropic low
mass clouds and both the shape of their mass function and
PDFs of other characteristics are strongly depended upon
parameters used for selection of the considered DM clouds.
In contrast, for population of richer clouds the PDFs of
cloud characteristics only weakly depend on the parameters
that select these clouds.

A strong difference between velocity dispersions $\sigma_U$
(\ref{sigu}) and $\sigma_i$ (\ref{fitv}) related to the motion
of clouds as a whole and to velocities within clouds is caused
by the large scale spatial modulation of the initial velocity
field in regions of the clouds formation. The same effect is
seen in estimates of the relative velocities of neighboring
clusters, $\Delta U$ (\ref{delt_u}).

\subsection{Angular momentum of the LSS elements.}

 Recently the angular momenta of the LSS elements are widely
discussed and compared with the alignments and angular momenta of
galaxies and their DM halos (see, e.g., Vitvitska et al. 2002;
Patiri et al. 2006; Lee\,\&\,Erdogdu 2007; Park et al. 2007;
Aragon--Calvo et al. 2007, 2008; Cuesta et al. 2008; Paz et al.
2008; Slosar et al. 2009; Jimenes et al. 2009; Zhang et al. 2009;
Lovell et al. 2010). The main inferences of the TTT and, in
particular, preferential orientation of the angular momenta along
the shortest axes of the LSS elements are basically confirmed.
Influence of the anisotropic matter infall into the LSS elements on
their angular momenta was discussed by Doroshkevich (1973) for
baryonic component and in Bernardo et al. (2002) for DM particles.
However, analysis of simulations shows that for DM component
contribution of this effect is moderate and our results obtained in
Sec. 3.4 are consistent with TTT predictions.

On the other hand, large scatter is found between the shape
of the host LSS element and the angular momenta of galaxies and
their DM halos incorporated in that element. This scatter is
expected and can be caused by action of several factors.

Thus, comparison of the functions $J\,\&\,j$ (\ref{jj}) shows
 that the process of  generation of angular
momentum of DM clouds is only moderately efficient (less than 20\%).
This fact in itself suggests a weak correlation between the angular
momenta of the host LSS element and high density low mass subclouds
and it basically explains the scatter mentioned above. Moreover,
more detailed analysis of the process of halo formation (Vitvitska
et al. 2002) shows strong evolution of the angular momentum of DM
halos and its dependence upon the angular momenta of merged
satellites. Perhaps, stronger correlation can be found between
properties of the host cloud and the system of DM halos and its
satellites.

Let us note also that this scatter is partly enhanced by
uncertainties in the determination of the complex shapes of the
LSS elements. This is specially important for observations when
the shape of such elements can be found only approximately.

\subsection{The DM LSS elements and the Ly--$\alpha$ forest.}

The main observed characteristics of the Ly-$\alpha$ forest
were analyzed in many publications (see, e.g. Demia\'nski et al.
2006 and references there). Detailed discussion
of the observed characteristics of the Ly-$\alpha$ absorbers
can be found, for example, in Kim, Cristiani \& D'Odorico (2002),
Kim et al. (2002, 2004).

There are some essential differences between the sample of
simulated DM clouds and the observed Ly-$\alpha$ absorbers.
The most important of them are:
\begin{enumerate}
\item{} The Ly--$\alpha$ forest is related with the LSS elements
        with sizes, masses and other parameters much
        smaller than the simulated ones.
\item{} The Ly--$\alpha$ forest is observed for larger
        neutral hydrogen column density, $N_{HI}\geq 10^{12}
        cm^{-2}$, which depends upon the UV background and is not
        directly linked with the DM component of absorbers.
\item{} The evolution of the observed Ly-$\alpha$ absorbers
        depends upon the poorly known UV background
    and can significantly differ from the evolution of the simulated
    DM LSS.
\end{enumerate}

These differences prevent a direct quantitative comparison of
characteristics of the simulated DM LSS and the observed
Ly--$\alpha$ forest. However, the absorbers are dominated by the DM
component, and, so, qualitative comparison of their properties and
evolution with the simulated one allows us to clarify some problems
of the forest evolution.

First of all the self similar character of evolution of the Doppler
parameter, $b$, is reflected in  the surprising stability of its PDFs
(Demia\'nski et al. 2006). The $b$ parameter strongly
depends upon the DM component and we can expect that the most
important aspects of its evolutions are closely correlated with the
characteristics of evolution of DM clouds.

The weak redshift dependence of the Doppler parameter and the
high degree of matter relaxation along the line of sight for
majority of the Ly-$\alpha$ absorbers are consistent with results
obtained in Sec. 3.3, 3.5 and 4.1. It is important also, that
the complex shape of the PDF $P(b)$ is quite similar to the PDF
$P(\sigma_3)$ plotted in Fig. \ref{fv}.

Our analysis in Sec. 4.2 cannot reproduce criteria of the
Ly-$\alpha$ forest selection and in particular the redshift
evolution of clumps separation (\ref{dsep}, \ref{dsep50}) does
not represent the evolution of linear number density of the
forest elements. The most important among the unknown factors are the
disintegration of richer forest elements into a system of high
density subclouds linked by extended low density bridges, and
variations of the ionizing UV background. Comparison of the LSS
evolution in simulations $S_{150}$ and $S_{50}$ confirms that
the small scale perturbations accelerate the disintegration of the
LSS elements and their internal structure becomes more complex.
These problems need to be studied in more details with more
representative and high resolution simulations.

\subsection*{Acknowledgments}
We wish to thank Dr. E. Tittley for many useful comments and
suggestions on this paper.
This paper was supported in part by the Russian Fund of Fundamental
Investigations grant Nr. 08-02-00159\,\&\, 09-02-12163, Federal
Program "Scientific and Pedagogical Personal Innovative of Russia"
Nr. 1336 and Polish Ministry of Science and Higher Education grant
NN202-091839.

\vfill\break

\end{document}